\documentclass{aa}

\usepackage{txfonts}
\usepackage{graphicx}
\usepackage{natbib}

\usepackage{multirow}

\newcommand{\chandra}{\emph{Chandra}}

\newcommand{\igrfourteen}{IGR\,J14003$-$6326}
\newcommand{\igreighteen}{IGR\,J18490$-$0000}
\newcommand{\hesssixteen}{HESS\,J1632$-$478}

\begin{document}


\title{A search for near infrared counterparts of 3 pulsar wind nebulae\thanks{Based on observations collected at the European Organisation for Astronomical Research in the Southern Hemisphere, Chile under ESO programs 080.D-0864, 081.D-0401, 084.D-0535 (P.I. Chaty)}}

\titlerunning{A search for near infrared counterparts of 3 pulsar wind nebulae}

  \author{
    P.A.~Curran\inst{1}
    \and S.~Chaty\inst{1}
    \and J.A.~Zurita Heras\inst{2}
    \and A.~Coleiro\inst{1}
  }


  
  \institute{
    Laboratoire AIM, CEA/IRFU-Universit\'e Paris Diderot-CNRS/INSU, CEA DSM/IRFU/SAp, Centre de Saclay, F-91191  Gif-sur-Yvette, France 
    \and Fran\c{c}ois Arago Centre, APC, Universit\'e Paris Diderot, CNRS/IN2P3, CEA/DSM, Observatoire de Paris, 13 rue Watt, 75205 Paris Cedex 13, France 
  }

\date{Received ; accepted}


\abstract
{While pulsar wind nebulae (PWNe) and their associated isolated pulsars are commonly detected at X-ray energies, they are much rarer at near infrared (nIR) and optical wavelengths.}
%
{Here  we examine three PWN systems in the Galactic plane  -- \igrfourteen, \hesssixteen\  and \igreighteen\ -- in a bid to identify optical/nIR emission associated with either the extended  PWNe or their previously detected  X-ray point sources.}
%
{We obtain optical/nIR images of the three fields with the ESO -- New Technology Telescope and apply standard photometric and astrometric calibrations.} 
%
{We find no evidence of any extended emission associated with the PWNe in any of the fields; neither do we find any new counterparts to the X-ray point sources, except to confirm the magnitude of the previously identified counterpart candidate of \igreighteen.}
%
{Further observations are required to confirm the association of the nIR source to \igreighteen\ and to detect counterparts to \igrfourteen\ and \hesssixteen, while a more accurate X-ray position is required to reduce the probability of a chance superposition in the field of the latter.}


\maketitle

\section{Introduction}\label{section:intro}

Along with the typical supernova remnant (SNR), a rapidly-rotating, highly-magnetised neutron star, or pulsar, is frequently an end product of a supernova explosion. The pulsar has particularly high levels of rotational energy which is dissipated via a highly relativistic particle wind. The interaction of this wind with the surrounding medium, i.e., the ejecta of the supernova explosion itself, causes a continuously refreshed shock wave known as a pulsar wind nebula (PWN). These PWNe emit via synchrotron and inverse Compton processes and are observed across the spectrum from radio to optical to X-ray and higher energies (for a detailed review of PWNe see, e.g. \citealt{Gaensler2006:ARA&A.44,Slane2011:heep.373}). The pulsars themselves, as well as being detected as point sources in radio and X-ray where the pulsations are observed, may be detected at optical or near infrared (nIR) wavelengths. However at these wavelengths the emission will be that of an isolated,  non-accreting  neutron star which is intrinsically dim and hence very few (12 out of $\sim 1800$) have been  detected in this regime \citep{Mignani2011:AdSpR.47}.

Here we examine three such PWN systems in the Galactic plane -- \igrfourteen, \hesssixteen\  and \igreighteen\ -- in a bid to identify extended optical/nIR (OIR) emission associated with the PWNe or optical/nIR counterparts to the X-ray point sources. The intrinsic OIR dimness of both PWNe and isolated neutron stars is further compounded in these cases by the high level of optical extinction in the direction of the Galactic plane \citep{schlegel1998:ApJ500}; hence observing at nIR wavelengths, where this is less pronounced, may increase the chance of a detection. However, at these positions in the Galactic plane nIR source density is relatively high so even with accurate X-ray positions care must be taken to understand the probability of chance superpositions. 
In section \ref{section:observations} we introduce our observations and reduction method while in section \ref{section:results}, after briefly introducing each source,  we detail the results of those observations.  
In section \ref{section:discussion} we discuss these results and summarise our findings in section \ref{section:conclusions}.
Throughout, positions (J2000) are given with 90\% confidence while all others values, including magnitudes, are given with $1\sigma$ confidence.


\section{ESO-NTT observations and data analysis}\label{section:observations}


 \begin{figure*} 
  \centering 
  \includegraphics[angle=-0]{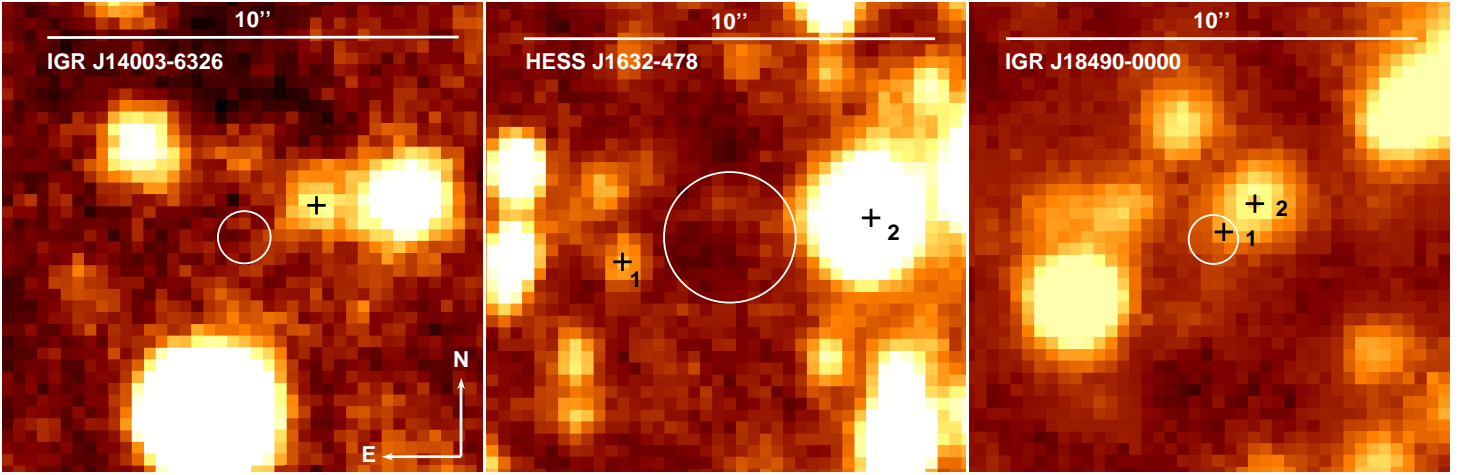}
  \caption{NTT-SofI $J$ (\igrfourteen) or $K_s$ (\hesssixteen, \igreighteen) band images of each of the three fields with the best X-ray positions (Table\,\ref{table:positions}; white circles) marked along with the positions of the optical sources (black crosses) discussed in section \ref{section:results}.} 
  \label{fig.Ks} 
\end{figure*}


Optical ($B, V, R, i$) and nIR ($J, H, K_s$) data were obtained with the ESO Faint Object Spectrograph and Camera (v.2; EFOSC2) and the Son of ISAAC (SofI) infrared spectrograph and imaging camera on the 3.58m ESO -- New Technology Telescope (NTT). Data were obtained on the nights of March 8, 2008 (\igreighteen), September 16 \& 18, 2008 (\igrfourteen; optical \& nIR) and March 27, 2010 (\hesssixteen). 
All data used a dithered pattern of $3\times60$\,seconds or $9\times 10$\,seconds per final image in the optical and nIR respectively (see Table\,\ref{table:observations}).

The NTT data were reduced using the {\small IRAF} package wherein crosstalk correction, flatfielding, sky subtraction, bias-subtraction and frame addition were carried out as necessary. 
The images were astrometrically calibrated against 2MASS \citep{Skrutskie2006:AJ.131} or USNO-B1.0 \citep{Monet2003:AJ125} within the GAIA package and quoted 90\% positional errors include a 0.16\arcsec\ 2MASS  systematic uncertainty.
Relative point spread function (PSF) photometry was carried out on the final images using the {\small DAOPHOT} package \citep{stetson1987:PASP99} within {\small IRAF}.  
Magnitudes were calibrated against \cite{Persson1998:AJ116} or \cite{Landolt1992:AJ.104} photometric standards, observed on the same night (Table\,\ref{table:positions}) and errors include both statistical and calibration errors. 
The equation, $i-I = (0.247 \pm 0.003) (R-I)$ \citep{Jordi2006A&A.460}, was used to transform the cataloged $I$ magnitudes of the standard stars into $i$ magnitudes with which to calibrate the images.

Images were inspected visually for any background emission that might be associated with the extended emission of the PWNe; this was done both to the reduced images and to those images after the PSF subtraction of point sources in the field.  We find no evidence of any extended emission associated with the PWN in any of the images, but due to uncertainties in the extent of the PWN and the quality of point source subtraction we are unable to put a flux limit on possible emission. Point source upper limits are approximated from the dimmest observable object in the field.

\begin{table}	
  \centering	
  \caption{Observation log.}
  \label{table:observations} 	
  \begin{tabular}{l l l l} 
    \hline\hline
    Field & Date  & & \\
    Filter & Exp (s) & Seeing (\arcsec) & Limit\\ 
    \hline 
    \igrfourteen & \multicolumn{3}{l}{September 16 \& 18, 2008} \\
    $Ks$& 90  & 1.7 &   $>17.2$   \\
    $H$	& 90  & 1.7 &   $>18.0$  \\
    $J$	& 90  & 1.7 &   $>19.2$  \\
    $i$	& 180 & 1.6 &   $>21.2$  \\
    $R$	& 180 & 1.8 &   $>21.0$  \\
    $V$	& 180 & 1.7 &   $>21.8$  \\
    $B$	& 180 & 1.7 &   $>22.3$ \\
    \hline 
    \hesssixteen & \multicolumn{3}{l}{March 27, 2010} \\
    $Ks$&  90 & 0.9 &  $>17.7$ \\
    \hline 
    \igreighteen & \multicolumn{3}{l}{March 8, 2008} \\
    $Ks$& 90 & 1.6 &  $>17.7$   \\
    \hline 
  \end{tabular}
\end{table}


\section{Results}\label{section:results}

\begin{table*}
  \centering	
  \caption{X-ray and optical positions (90\% uncertainties) and optical apparent magnitudes for the three X-ray point sources (first lines) and the nearby optical sources (numbered lines), as labelled in Figure \ref{fig.Ks}. } 	
  \label{table:positions} 	
  \begin{tabular}{r l l l l} 
    \hline\hline
    \multicolumn{1}{l}{Source}   & & & \\
    (optical source) & RA & Declination & Error (\arcsec) & Magnitudes\\ 
    \hline 
    \multicolumn{1}{l}{\igrfourteen$^a$}  & 14:00:45.69  & $-$63:25:42.6 & 0.64  & \\
    (1) & 14:00:45.45  & $-$63:25:41.8 & 0.3 & $J = 18.6 \pm 0.2$, $i = 20.90 \pm 0.15$\\
    \hline 
    \multicolumn{1}{l}{\hesssixteen$^b$} & 16:32:08.8 & $-$47:49:01 & 1.6 & \\
    (1) & 16:32:09.06 & $-$47:49:01.6 & 0.3  & $K_S = 16.74 \pm 0.05$ \\
    (2) & 16:32:08.47 & $-$47:49:00.6 & 0.16   & $K_S = 11.35 \pm 0.01$, $K^{\dagger} = 11.695 \pm 0.040, H^{\dagger} = 13.284 \pm 0.065$  \\
    \hline 
    \multicolumn{1}{l}{\igreighteen$^c$} & 18:49:01.59 & $-$00:01:17.73 & 0.6 &  \\
    (1) & 18:49:01.57 & $-$00:01:17.6 & 0.3   &  $K_S = 17.2 \pm 0.4$  \\
    (2) & 18:49:01.52 & $-$00:01:16.9 & 0.2   &  $K_S = 15.58 \pm 0.05$   \\
    \hline 
  \end{tabular}
  \begin{list}{}{}
  \item[] References to X-ray positions: $^a$\cite{Tomsick2009:ApJ701}, $^b$\cite{Balbo2010:A&A.520}, $^c$\cite{Ratti2010:MNRAS.408}
  \item[] $^{\dagger}$ 2MASS magnitudes
  \end{list}
\end{table*}

We find no evidence of any extended emission associated with the PWN in any of the images, even after the PSF subtraction of point sources in the field. 
Neither do we find any  nIR counterparts to the X-ray point sources, except to confirm the $K_s$ magnitude of the previously identified counterpart of \igreighteen\ \citep{Ratti2010:MNRAS.408}.  \\

\subsection{\igrfourteen}

Based on \chandra\ spectra, \cite{Tomsick2009:ApJ701} confirm that  \igrfourteen\ \citep{Keek2006:ATel.810}  is an SNR with a PWN, while \cite{Renaud2010:ApJ.716} discovered a 31.18\,ms X-ray/radio pulsar at its centre. \citeauthor{Tomsick2009:ApJ701} do not find any higher energy (TeV, GeV) counterparts to the source but  \citeauthor{Renaud2010:ApJ.716} obtain radio observations which reveal counterparts to both the point source and the PWN.

We find no source within $3\sigma$ of the \chandra\ poition of \igrfourteen, down to the magnitudes given in Table\,\ref{table:observations}. However there is a dim source detected at  $J = 18.6 \pm 0.2$ and marginally at $i = 20.90 \pm 0.15$, at RA, Dec = 14:00:45.45, $-$63:25:41.8 ($\pm 0.3$\arcsec) or $\sim 5\sigma$ from the \chandra\ position. We are unable to discern if this is a point or extended source due to its faintness but it is likely unrelated to the X-ray source given the distance discrepancy.

\subsection{\hesssixteen}

\hesssixteen\ was initially suggested to be associated with IGR\,J16320-4751 \citep{Aharonian2006:ApJ.636} 
though this was subsequently rejected after deep XMM-\emph{Newton} observations of the source \citep{Balbo2010:A&A.520}. It was instead associated with an independent X-ray point source with diffuse emission. 
These authors also obtained data from radio and high energy  archives or catalogs to describe the nature of the source, which they suggest is an energetic PWN with a, yet to be confirmed, central pulsar.

We find no source within the XMM error circle of \hesssixteen, down to a limiting magnitude of $K_s > 17.7$. However, there is a dim source 
2.6\arcsec $\sim 2.6 \sigma$ to the East (source 1) with  a magnitude of  $K_S = 16.74 \pm 0.05$, though we cannot propose this as the counterpart with any certainty given the lack of colour information and weak positional agreement. Neither can we compare the colour of this source to other sources in the  field  to demonstrate that it has similar properties and is thus likely to be a field source, unrelated to the suggested pulsar. There is also a bright 2MASS  source (16320846$-$4749005) to the West (source 2), though it is at a distance of $\sim 3.5\sigma$ so is even less likely related to the high energy source.

\subsection{\igreighteen}

First reported by \cite{Molkov2004:AstL.30}, \igreighteen\  is a PWN (e.g., \citealt{Ratti2010:MNRAS.408}) with a confirmed 38.5\,ms pulsar \citep{Gotthelf2011:ApJ729}. \citeauthor{Ratti2010:MNRAS.408} also suggest a nIR counterpart of magnitude $K_S = 16.4 \pm 0.1$ at RA, Dec = 18:49:01.563, $-$00:01:17.35 ($\pm 0.1$\arcsec) but find no evidence of any nIR extension as one might expect for a PWN.

Within the \chandra\ error circle of \igreighteen, we confirm the proposed nIR counterpart of \cite{Ratti2010:MNRAS.408} at a consistent magnitude of $K_s = 17.2 \pm 0.4$ (source 1). We also note, as those authors did, that the object is heavily blended with a nearby source of magnitude $K_s = 15.58 \pm 0.05$  (source 2). 
Due to the lack of colour information for the proposed counterpart we are unable to gain any information regarding its equivalent spectral classification, or to compare it to  other field sources to demonstrate that it has different properties that may indicate  it is associated with the  pulsar.


\section{Discussion}\label{section:discussion}

The lack of detected compact counterparts for \igrfourteen\ and \hesssixteen, particularly in the optical, is not surprising given the high levels of Galactic extinction \citep{schlegel1998:ApJ500} towards the sources:  $E_{B-V} = 3.83$  ($A_K \sim 1.3$, $A_V \sim 12$; \citealt{cardelli1989:ApJ345}) and $E_{B-V} = 11.18$ ($A_K \sim 3.9$) respectively. 
Likewise \igreighteen\ suffers significant extinction of $E_{B-V} = 6.62$ ($A_K \sim 2.3$).  Note that all these extinctions should be treated with caution as estimates so close to the Galactic plane ($<5\deg$) are unreliable. 
The magnitude limits of the non-detections are consistent with the $V$ band detections of other isolated neutron stars \citep{Mignani2011:AdSpR.47}; excluding the Crab at $V=16.6$, these range from 22 to 28 magnitudes (corresponding to $K_s \gtrsim 18.6$ for a spectral index of $-1$), at low optical extinctions ($E_{B-V} \lesssim 0.2$) in the nearby Galaxy ($\lesssim 1$kpc for most). On the other hand, the detection of a proposed counterpart to \igreighteen\ at $K_s \sim 17$ is significantly brighter than any other optical identification of an isolated neutron star, except for the Crab; it should also be noted that the proposed counterpart is significantly brighter than a simple power-law extrapolation of the X-ray spectra  \citep{Gotthelf2011:ApJ729} which implies a magnitude of $K_s \sim 23$ (uncorrected for Galactic extinction). While the power-law extrapolation of the X-ray spectra of \hesssixteen\  \citep{Balbo2010:A&A.520} is not constraining, that of  \igrfourteen\ \citep{Renaud2010:ApJ.716}  implies that a counterpart should be much dimmer than the observed optical limit, at approximately  $K_s \sim 22$. 
The extrapolations should however be treated with caution as the extracted X-ray spectra themselves may suffer from contamination from the surrounding PWN and thus inaccurate spectral slopes and fluxes. 
By using a simple power-law extrapolation, we have assumed that the spectra do not evolve between the X-ray and nIR regimes, while it quite possibly breaks to a shallower spectral index or, alternatively, may be described by thermal emission which naturally turns over at lower frequencies. 
We have also implicitly assumed that the nIR emission originates from the same emission region as the X-ray spectra, which is not necessarily true, but assuming that it is,  the above approximations can be treated as lower magnitude limits on how bright we might expect a nIR counterpart to be, before correction for Galactic extinction.

To approximate the probability of a chance superposition of the 90\% X-ray positions of the compact sources with random sources in the respective fields, we calculate $P \approx 1-\exp^{-(\rho_N \times A_{Err})}$; where $\rho_N$ is the surface area number density of observed sources down to the limiting magnitude and $A_{Err}$ is the area of the X-ray positional error. 
For \igrfourteen\ and \igreighteen, which have positions accurate at the sub-arcsecond level, we find probabilities of chance superpositions of 3\% and 5\% respectively, making the positional coincidence of \igreighteen\ with a nIR object reasonably compelling evidence for its association.  However, for \hesssixteen, with only an arcsecond accurate position, the probability is much greater at 50\%, though the source is in a relatively less densely populated region of the field, where the local probability is $\sim 40\%$. In this case, even if a source had been found within the 90\% error circle, its association with the X-ray source would be weak, meaning source 1 at a distance of $\sim 2.6\sigma$ is likely unrelated. To associate a nIR counterpart with this source based on positional coincidence will require a significantly better constrained position from e.g. \chandra.

Along with the optical extinction in their directions, the detection of extended emission from the PWNe is complicated by the high level of background in nIR observations as well as the generally high density of field sources in the Galactic plane which contaminate the background. Deep nIR images from larger telescopes better able to resolve field sources, may be able to detect PWN emission, as well as increasing the probability of detecting the compact sources, but in these directions source confusion will always be a major impediment to detection of point or extended sources.

\section{Conclusions}\label{section:conclusions}

We find no evidence of any extended nIR emission associated with the PWNe in any of the fields or any new counterparts to the X-ray point sources, but we do confirm the magnitude of the previously suggested counterpart of \igreighteen. While there is a low probability of chance coincidence of the X-ray position with a nIR object in this field, the candidate source is significantly brighter than most other isolated pulsars and brighter than a simple extrapolation of the X-ray spectra to the nIR, making its association with the X-ray source less certain. If future observations can confirm the association it seems that an additional emission component will be necessary to explain the excess nIR flux. 
The non-detection of the other two sources, \igrfourteen\ and  \hesssixteen, may be understood by the high level of Galactic extinction in their direction and by the intrinsically faint nature of isolated neutron stars, which the compact objects are assumed to be; it may be reasonable to expect the counterparts of these sources to be at magnitudes  $K_s \gtrsim 18.6$.


\begin{acknowledgements}   
We thank the referee for their constructive comments. 
This work was supported by the Centre National d'Etudes Spatiales (CNES) and is based on observations obtained with MINE: the Multi-wavelength INTEGRAL NEtwork.
This work made use of NASA's Astrophysics Data System. 
IRAF is distributed by the National Optical Astronomy Observatory, which is operated by the Association of Universities for Research in Astronomy (AURA) under cooperative agreement with the National Science Foundation.
GAIA was created by the now closed Starlink UK project funded by the Particle Physics and Astronomy Research Council (PPARC), and has been more recently supported by the Joint Astronomy Centre Hawaii funded by PPARC and now its successor organisation the Science and Technology Facilities Council (STFC). 
\end{acknowledgements}

\end{document}